\documentclass{article}

\PassOptionsToPackage{sort&compress,numbers,super}{natbib}
\usepackage[preprint]{neurips_2024}

\usepackage[utf8]{inputenc} 
\usepackage[T1]{fontenc}    
\usepackage{hyperref}       
\usepackage{url}            
\usepackage{booktabs}       
\usepackage{amsfonts}       
\usepackage{nicefrac}       
\usepackage{microtype}      
\usepackage{xcolor}         
\usepackage{graphicx}
\usepackage{booktabs}
\usepackage{amsmath}
\usepackage[numbers,sort&compress]{natbib}

\title{Directly Optimizing for Synthesizability in Generative Molecular Design using Retrosynthesis Models}

\author{
  Jeff Guo\textsuperscript{1,2}, Philippe Schwaller\textsuperscript{1,2} \\
  \textsuperscript{1}\'Ecole Polytechnique F\'{e}d\'{e}rale de Lausanne (EPFL) \\
  \textsuperscript{2}National Centre of Competence in Research (NCCR) Catalysis \\
  \texttt{\{jeff.guo,philippe.schwaller\}@epfl.ch} \\
}

\begin{document}

\maketitle

\begin{abstract}
    Synthesizability in generative molecular design remains a pressing challenge. Existing methods to assess synthesizability span heuristics-based methods, retrosynthesis models, and synthesizability-constrained molecular generation. The latter has become increasingly prevalent and proceeds by defining a set of permitted actions a model can take when generating molecules, such that all generations are anchored in "synthetically-feasible" chemical transformations. To date, retrosynthesis models have been mostly used as a post-hoc filtering tool as their inference cost remains prohibitive to use directly in an optimization loop. In this work, we show that with a sufficiently sample-efficient generative model, it is straightforward to directly optimize for synthesizability using retrosynthesis models in goal-directed generation. Under a heavily-constrained computational budget, our model can generate molecules satisfying a multi-parameter drug discovery optimization task while being synthesizable, as deemed by the retrosynthesis model. The code is available at \url{https://github.com/schwallergroup/saturn/tree/synth}.
\end{abstract}

\section{Introduction}

Generative molecular design for drug discovery has recently seen a surge of experimental validation, with many candidate molecules progressing into clinical trials~\cite{generative-design-review}. However, the synthesizability of generated designs remains a pressing challenge. Regardless of how "good" generated molecules are, they must be synthesized and experimentally validated to be of use, and work has shown that many generative models propose molecules for which finding a viable synthetic route for, is \textit{at the very least} not straightforward~\cite{gao2020synthesizability, fake-it-until-you-make-it}. Existing works tackle synthesizability in generative molecular design either by heuristics~\cite{sa-score, syba}, learning synthetic complexity from reaction corpus~\cite{sc-score}, retrosynthesis models which predict synthetic routes~\cite{lstm-retro, segler2017neural, coley2017computer, segler-retro-mcts, aizynthfinder-dataset, aizynthfinder, aizynthfinder-4, ra-score, askcos, ibm-rxn, retrognn, desp}, or enforce a notion of synthesizability directly in the generative process~\cite{synopsis, dogs, molecule-chef, barking-up-the-right-tree, chembo, pgfs, reactor, synnet, synthemol, synflownet, rgfn}.

Recently, synthesizability-constrained generative models~\cite{synopsis, dogs, molecule-chef, barking-up-the-right-tree, chembo, pgfs, reactor, synnet, synthemol, synflownet, rgfn} have become increasingly prevalent. A typical metric to \textit{quantify} synthesizability is whether a retrosynthesis model can solve a route for the generated molecules~\cite{synflownet, rgfn}. It is common practice to apply retrosynthesis models during post-hoc filtering due to their inference cost~\cite{gao2020synthesizability, fake-it-until-you-make-it}.

On the other hand, sample efficiency is also a pressing challenge, which concerns with how many oracle calls (computational predictions of molecular properties) are required to optimize an objective function. When these oracle calls are computationally expensive, such as in binding affinity predictions, there is a practical limit to an \textit{acceptable} oracle budget for real-world model deployment. The Practical Molecular Optimization (PMO) benchmark~\cite{pmo} highlighted the importance of sample efficiency and since then, more recent works have explicitly considered an oracle budget~\cite{freed, fu2022reinforced, mood, geam, tacogfn, linkinvent, reinvent-al, augmented-memory, beam-enumeration, saturn}.

\begin{figure}[!ht]
\centering
\includegraphics[width=1.0\columnwidth]{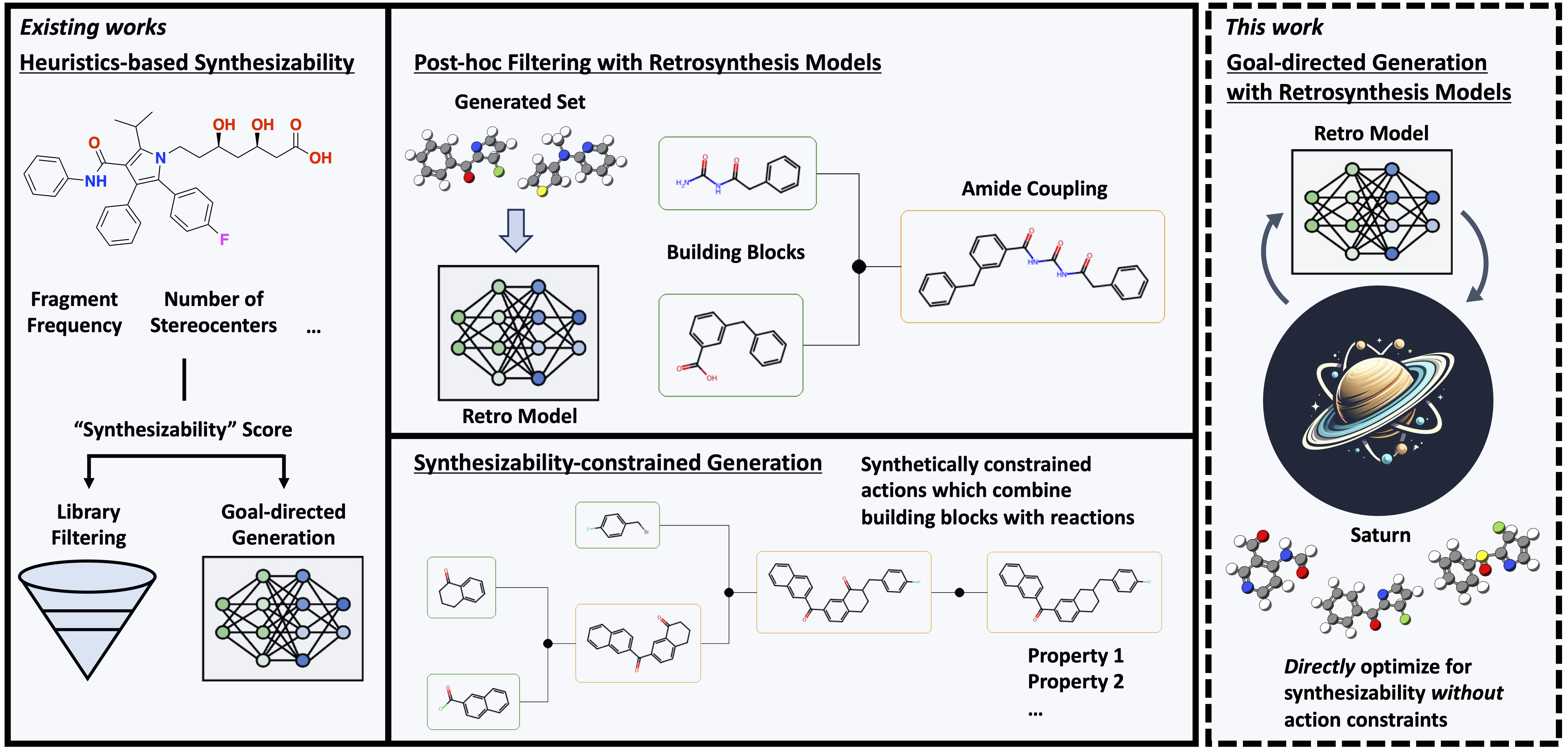} 
\caption{Overview of algorithmic methods to handle synthesizability in generative molecular design.}
\label{fig:overview}
\end{figure}

Recently, Saturn~\cite{saturn}, which is a language-based molecular generative model leveraging the Mamba~\cite{mamba} architecture, displayed state-of-the-art sample efficiency compared to 22 models. In this work, we build on Saturn and show that with a \textit{sufficiently sample-efficient} model, one can treat retrosynthesis models as an oracle~\cite{ekborg2024novo} and directly optimize for generating molecules where synthesis routes can be solved for (Fig. \ref{fig:overview}). We compare to the recent Reaction-GFlowNet (RGFN)~\cite{rgfn} model and show that Saturn can optimize their proposed multi-parameter optimization (MPO) task to generate molecules with good docking scores (to predict binding affinity) and is synthesizable (as deemed by a retrosynthesis model) with 1/400th the oracle budget (1,000 calls instead of 400,000).

\textbf{We strictly emphasize that we neither claim to solve synthesizability nor claim our model \textit{guarantees} synthesizability.} Rather, the take-home message of this work is that if one wants to optimize certain properties, then they should be included in the MPO objective function. If the downstream metric is whether a retrosynthesis tool can solve a route for the generated molecules~\cite{gao2020synthesizability, synflownet, rgfn}, then the tool itself should be part of the objective function (given that the model is not synthesizability-constrained).

\section{Related Work}

\textbf{Synthesizability Metrics.} Quantifying and defining synthesizability is non-trivial and early metrics assess \textit{molecular complexity} rather than synthesizability explicitly. Exemplary works include the Synthetic Accessibility (SA) score~\cite{sa-score} and SYnthetic Bayesian Accessibility (SYBA)~\cite{syba} which are based on the frequency of chemical groups in databases. The Synthetic Complexity (SC) score~\cite{sc-score} is trained on Reaxys data to measure molecular complexity and implicitly considers the number of synthetic steps required to make a target molecule. There is a correlation between these scores and whether retrosynthesis tools can solve a route~\cite{synthesizability-metrics-assessment}. The recent Focused Synthesizability (FS) score~\cite{fs-score} incorporated domain-expert \textit{preferences}~\cite{molskill} to assess synthesizability.

\textbf{Retrosynthesis Models.} Given a target molecule, retrosynthesis models propose viable synthetic routes by combining commercial building blocks (starting reagents) with reaction templates (coded patterns that map chemical reaction compatibility) or template-free approaches (learned patterns from data). Exemplary examples include the first work applying Monte Carlo tree search (MCTS) for retrosynthesis~\cite{segler-retro-mcts}, SYNTHIA~\cite{chematica-1, chematica-2}, AiZynthFinder~\cite{aizynthfinder-dataset, aizynthfinder, aizynthfinder-4}, ASKCOS~\cite{askcos}, Eli Lilly's LillyMol retrosynthesis model~\cite{eli-lilly-retro}, Molecule.one's M1 platform~\cite{molecule-one}, and IBM RXN~\cite{ibm-rxn, schwaller2020predicting, thakkar2023unbiasing}. We further highlight surrogate models including Retrosynthesis Accessibility (RA) score~\cite{ra-score} and RetroGNN~\cite{retrognn} trained on the output of retrosynthesis models for faster inference. Note that these models output a score rather than synthetic routes.

\textbf{Synthesizability-constrained Molecular Generation.} More recently, molecular generative models have been designed with a notion of synthesizability, for example by enforcing transformations from a set of permitted reaction templates. Expansion methods include SYNOPSIS~\cite{synopsis}, Design of Genuine Structures (DOGS)~\cite{dogs}, and RENATE~\cite{renate}. Other models include MOLECULE CHEF~\cite{molecule-chef}, Synthesis Directed Acyclic Graph (DAG)~\cite{barking-up-the-right-tree}, ChemBO~\cite{chembo}, SynNet~\cite{synnet}, and SyntheMol~\cite{synthemol}. Models that also use reinforcement learning (RL) include Policy Gradient for Forward Synthesis (PGFS)~\cite{pgfs}, Reaction-driven Objective Reinforcement (REACTOR)~\cite{reactor}, and LibINVENT~\cite{libinvent}. Recent works have equipped GFlowNets~\cite{gflownets-foundation} with reaction templates, including SynFlowNet~\cite{synflownet} and RGFN~\cite{rgfn}. Finally, very recent work proposes a new paradigm of "projecting" unsynthesizable molecules into similar, but synthesizable analogs~\cite{chem-projector}.

\textbf{Goal-directed Generation with Synthesizability Metrics.} An alternative to synthesizability-constrained molecular generation is to task molecular generative models to also optimize for synthesizability metrics~\cite{fs-score}, with common ones being SA score~\cite{gao2020synthesizability, fake-it-until-you-make-it}. Although SA score assesses molecular complexity, it is correlated with whether AiZynthFinder can solve a route~\cite{synthesizability-metrics-assessment}. Generally, more confidence is placed on the output of retrosynthesis models in assessing synthesizability and this is reflected in works that assess model performance on whether generated molecules have a solved route~\cite{synflownet, rgfn, chem-projector}. In this work, we propose to directly incorporate retrosynthesis tools as an oracle in the MPO objective function and show that generated molecules satisfy MPO objectives.

We end this section by reinforcing that quantifying synthesizability is non-trivial and neither reaction templates nor retrosynthesis tools \textit{guarantee} synthesizability. Notably, reaction templates depend on the granularity of their definition, for instance with the inclusion or omission of incompatibilities which affects the false positive rate of matching reagents~\cite{imperfect-reaction-templates, chematica-natural-products}. A concrete example of this is the original paper reporting Enamine REAL which is a "make-on-demand" commercial database with a stated \textasciitilde{80\%} synthesis success rate~\cite{enamine-real}. Recently, SyntheMol~\cite{synthemol} which enforces reaction templates during molecular generation, ordered 70 compounds from Enamine REAL with 58 successful syntheses (\textasciitilde{83\%}). We wish to emphasize that the point of drawing attention to this is strictly to support our statement that neither reaction templates, make-on-demand libraries (often generated by reaction templates), nor retrosynthesis tools \textit{guarantee} synthesizability. "Make-on-demand libraries" are a remarkable resource.

\section{Methods}

In this section, we describe in detail the experimental design and highlight caveats in the results. Firstly, we use Saturn~\cite{saturn} as the generative model which uses RL for goal-directed generation and has high sample efficiency. For details of the model, we refer to the original work~\cite{saturn}. In Saturn, we newly implement AiZynthFinder~\cite{aizynthfinder-dataset, aizynthfinder, aizynthfinder-4} and QuickVina2-GPU-2.1~\cite{autodockvina, quickvina2, quickvina2-gpu-2.1} (for docking) as oracles to match the case study in RGFN~\cite{rgfn}.

The RGFN work assesses the synthesizability of generated molecules using the quantitative estimate of drug-likeness (QED)~\cite{qed}, SA score~\cite{sa-score}, and whether AiZynthFinder~\cite{aizynthfinder-dataset, aizynthfinder, aizynthfinder-4} can solve a route. The authors state that the latter better estimates synthesizability. This statement is supported by recent work highlighting how AiZynthFinder predictions can augment medicinal chemists' decision-making, which led to real-world impact in commercial drug discovery projects~\cite{az-aizynthfinder}. The RGFN work features three case studies where the objective function is either to optimize a proxy model for docking scores, optimize a proxy model for biological activity classification, or optimize QuickVina2-GPU-2.1~\cite{autodockvina, quickvina2, quickvina2-gpu-2.1} docking scores directly. We choose to compare our model on the latter task because proxy models, while offering faster inference, suffer from domain out-of-applicability if generated molecules deviate too far from the training data. This was also stated by the authors and was the motivation for designing the docking case study~\cite{rgfn}.

\textbf{Experimental Caveats.} As the code for RGFN~\cite{rgfn} is not released, we implement their oracle function ourselves. In Appendix \ref{appendix:rgfn-oracle}, we describe the steps we took to reproduce their case study faithfully. Here, we instead highlight caveats that make the comparison not exactly apples to apples for transparency.

\begin{enumerate}
    \item{\textbf{Pre-training:} Saturn is pre-trained with either ChEMBL 33~\cite{chembl} or ZINC~\cite{zinc250k}. These datasets, containing bio-active molecules, inherently bias the learned distribution to already known synthesizable entities~\cite{gao2020synthesizability}. On the other hand, RGFN defines a state space based on reaction templates and building blocks. We note, however, that these are common pre-training datasets that many generative models in literature are pre-trained with.}
    \item{\textbf{Quantifying Synthesizability:} RGFN handles synthesizability by combining building blocks with reaction templates. By contrast, Saturn is handling synthesizability by optimizing AiZynthFinder. It could be the case that RGFN's reaction templates represent "true" synthesizability better in some cases. We note, however, that RGFN also evaluates synthesizability using AiZynthFinder, and in principle, any building blocks and templates used in RGFN could be added to a retrosynthesis model.
    \item{\textbf{Docking Results Filtering}.} RGFN filters the best generated molecules by whether they pass PoseBusters~\cite{posebusters} checks (plausible physicality). We do not consider this as this is essentially an artefact of the oracle. Provided a more accurate oracle, failure naturally decreases. We note that QuickVina2-GPU-2.1 outputs almost all pass the PoseBusters checks (see Appendix H in the RGFN~\cite{rgfn} work). Subsequently, RGFN filters molecules to be dissimilar (< 0.4 Tanimoto similarity) to known aggregators based on the Aggregation Advisor dataset~\cite{ucsf-aggregation} as the \textit{final set}. We also do not consider this. RGFN reports statistics \textit{before} this final aggregator filtering and \textit{these} are the results we compare to (Table 1 in the RGFN~\cite{rgfn} work).}
    \item{\textbf{Objective Function.}} The RGFN work (which also reports results for GraphGA~\cite{graphga}, SyntheMol~\cite{synthemol}, and FGFN~\cite{bengio2021flow}), defines the objective function to only optimize for docking score, but assesses generated molecules also by their QED and SA scores. It is unclear the performance of these models if the objective function were modified to also enforce these properties.
\end{enumerate}

Still, despite these caveats, the message we convey is that if one wants to optimize for downstream metrics, then they should be included in the MPO objective function. This is often impractical because certain oracles are computationally expensive and generative models are not efficient enough to directly optimize them. Provided a model \textit{is} sufficiently sample-efficient, generative models can be tasked to optimize \textit{anything} (this does \textit{not} mean that it will \textit{always} be able to optimize the objective under the budget).

\textbf{Experimental Setup.} Following the RGFN~\cite{rgfn} work, the case study is to generate synthesizable molecules with good docking scores (using QuickVina2-GPU-2.1~\cite{autodockvina, quickvina2, quickvina2-gpu-2.1}) to ATP-dependent Clp protease proteolytic subunit (ClpP). The objective function is:

\begin{equation}
R_{RGFN}(x) = Docking\ Score(x)
\label{eq:rgfn-reward}
\end{equation}

where $x$ is a generated molecule. In Saturn, we apply reward shaping so that $R_{RGFN}(x) \in [0, 1]$. As the purpose of this short paper is to convey that retrosynthesis models can be directly optimized as an oracle, we further define two objective functions:

\begin{equation}
\small
R_{\text{All MPO}}(x) = \left( \text{Docking Score}(x) \times \text{QED}(x) \times \text{SA Score}(x) \times \text{AiZynthFinder}(x) \right)^{\frac{1}{4}} \in [0, 1]
\label{eq:all-mpo-reward}
\end{equation}

\begin{equation}
R_{Double\ {MPO}}(x) = \left( Docking\ Score(x) \times AiZynthFinder(x) \right)^{\frac{1}{2}} \in [0, 1]
\label{eq:double-mpo-reward}
\end{equation}

See Appendix \ref{appendix:reward-shaping} for reward shaping details to normalize Eq. \ref{eq:all-mpo-reward} and \ref{eq:double-mpo-reward} $\in$ [0, 1] and the exponential term which is from the product aggregator that outputs the final reward. The rationale for $R_{All\ {MPO}}$ (Eq. \ref{eq:all-mpo-reward}) is because RGFN evaluates generated molecules also by their QED, SA score, and whether AiZynthFinder can solve a route. Since these are the downstream metrics, we include them in the objective function. The rationale for $R_{Double\ {MPO}}$ is to illustrate a contrast in optimization difficulty as $R_{All\ {MPO}}$ is inherently more challenging. Still, we show in the Results section that both objective functions can be optimized. 

All Saturn experiments are run across 10 seeds (0-9 inclusive) with 1,000 oracle calls. We note this is 1/400th of the oracle budget of the RGFN work (400,000 calls). We compare with RGFN and also GraphGA~\cite{graphga}, SyntheMol~\cite{synthemol}, and Fragment-based GFlowNet (FGFN)~\cite{bengio2021flow}. We do not run these models ourselves and take the results from the RGFN work.

\textbf{Metrics.} Following the RGFN~\cite{rgfn} work, a \textbf{Mode} is defined as a molecule with docking score < -10. \textbf{Discovered Modes} denotes the set of generated Modes that also possess Tanimoto similarity < 0.5 to every other mode. We note that Modes with > 0.5 Tanimoto similarity with other Modes are still valuable, as given a pair of "similar" molecules, there can be a clear preference if for example, one of the molecules contains an undesired substructure. For Saturn results, we additionally report \textbf{Yield} which denotes the total number of unique molecules generated with docking score < -10.

\section{Results and Discussion}
We devise three experiments: optimizing only docking score (following the RGFN~\cite{rgfn} work), jointly optimizing docking and AiZynthFinder, and lastly, showing that AiZynthFinder can \textit{still} be directly optimized as an oracle even if none of the molecules in the training data for the generative model can be solved by AiZynthFinder. We construct a custom dataset for this last case study.

\subsection{Experiment 1: Optimizing only docking score leads to unreasonable molecules}

\begin{figure}[!ht]
\centering
\includegraphics[width=1.0\columnwidth]{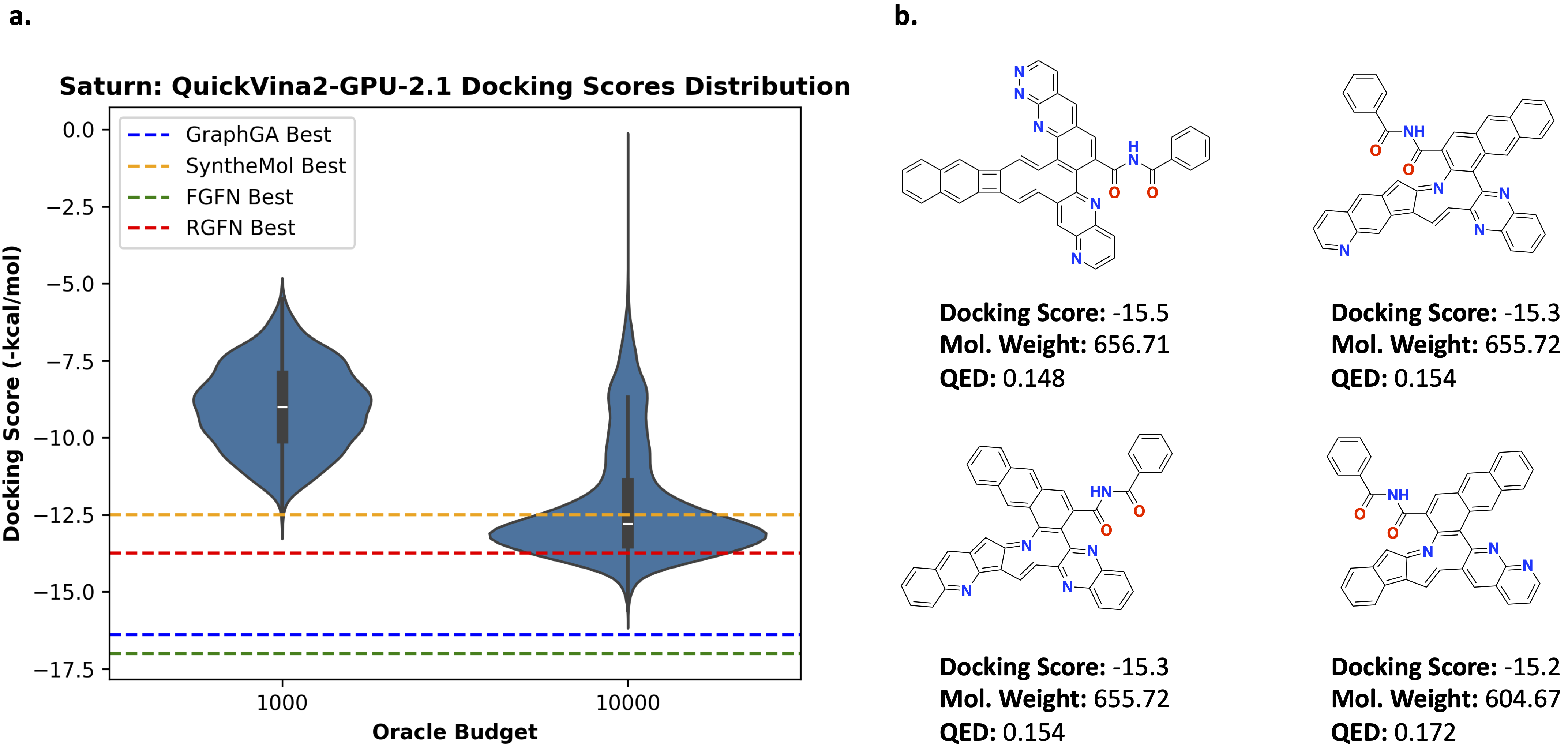} 
\caption{Experiment 1: Optimizing only docking scores. \textbf{a.} Distribution of docking scores at varying oracle budgets. The best docking score across comparison methods (taken from the RGFN~\cite{rgfn} work) are annotated as dotted lines. \textbf{b.} Example lipophilic molecules generated by Saturn with the best docking scores.}
\label{fig:greasy}
\end{figure}

We first present results for the $R_{RGFN}$ (Eq. \ref{eq:rgfn-reward}) objective function which only optimizes for docking scores against ClpP. It is generally not advised to optimize this in isolation because docking oracles can be highly exploitable, such that lipophilic (lots of carbon atoms and high logP) molecules (promiscuous binders with solubility issues~\cite{greasy-docking}) receive good docking scores. We show that with 10,000 oracle calls, Saturn (trained on ChEMBL 33~\cite{chembl}) generates molecules with approximately the same best QuickVina2-GPU-2.1~\cite{autodockvina, quickvina2, quickvina2-gpu-2.1} docking scores compared to GraphGA~\cite{graphga}, SyntheMol~\cite{synthemol}, FGFN~\cite{bengio2021flow}, and RGFN~\cite{rgfn} which were run with 400,000 oracle calls (40x higher budget). We perform one replicate here as we only want to convey that the objective function is highly exploitable. Fig. \ref{fig:greasy}a shows the distribution of docking scores at varying oracle budgets. We illustrate how the docking oracle can be exploited in Fig. \ref{fig:greasy}b which shows the best molecules generated by Saturn. Although possessing good docking scores, they are lipophilic with high molecular weight and low QED. Consequently, these are not meaningful molecules. Table 1 in the RGFN~\cite{rgfn} work shows that the best generated molecules across various models also have low QED: GraphGA (\textasciitilde{0.32}), FGFN (\textasciitilde{0.22}), and RGFN (\textasciitilde{0.23}), suggesting that they are also exploiting the docking oracle. We note that SyntheMol has slightly higher QED (\textasciitilde{0.45}).

\subsection{Experiment 2: Directly optimizing for synthesizability using AiZynthFinder}

\begin{table}[ht]
\centering
\scriptsize
\caption{Synthesizability metrics for top-k Modes (\textbf{molecules with docking score < -10}). Results are taken from the RGFN~\cite{rgfn} paper (it was not stated how many replicates the models were run for). Mol. weight, QED, and SA score results are for the top-500 Modes. AiZynth results are for the top-100 Modes. NR denotes "not reported". All Saturn experiments were run across 10 seeds (0-9 inclusive). The mean and standard deviation are reported. Both Yield and Modes are reported. The number after the configuration denotes the number of successful replicates out of 10 (Modes $\geq$ 1). For Saturn, none of the configurations found 100 Modes in 1,000 oracle calls so the metrics are reported for however many Modes were found. \\ $^{a}$ Less than 400,000 oracle calls as SyntheMol~\cite{synthemol} roll-outs took time. The RGFN authors decided to match the wall time instead.}
\label{table:synth-metrics}
\begin{tabular}{@{}lcccccc@{}}
\toprule
Method    & Modes (Yield) & Mol. weight (↓) & QED (↑) & SA score (↓) & AiZynth (↑) & Oracle calls \\ 
          &       &                 &         &             &             & (Wall time) \\ 
\midrule
\textbf{Previous work} & & top-500 & top-500 & top-500 & top-100 & \\ 
\midrule
GraphGA~\cite{graphga}   & NR     & 521.0 ± 31.8    & 0.32 ± 0.07 & 4.14 ± 0.51 & 0.00 & 400,000 (NR)            \\[3pt]
SyntheMol~\cite{synthemol} & NR     & 458.2 ± 60.7    & 0.45 ± 0.16 & 2.86 ± 0.56 & 0.56 & 100,000$^{a}$ (72h)   \\[3pt]
FGFN~\cite{bengio2021flow}  &   NR     & 548.6 ± 42.9    & 0.22 ± 0.03 & 2.94 ± 0.54 & 0.25 & 400,000 (NR)      \\[3pt]
RGFN~\cite{rgfn}      & NR    & 526.2 ± 37.6    & 0.23 ± 0.04 & 2.83 ± 0.22 & 0.65 & 400,000 (72h)              \\[3pt] 
\midrule
\textbf{$R_{All\ {MPO}}$} \textbf{(ours)} & \multicolumn{6}{c}{4 objectives (Docking, QED, SA, AiZynth)} \\
\midrule
Saturn-ChEMBL (10)   & 4 ± 1 (5 ± 3)    & 367.7 ± 15.7    & 0.70 ± 0.13 & 2.11 ± 0.19 & 0.91 ± 0.11  & 1,000 (2.9h ± 34m)            \\[3pt]
Saturn-GA-ChEMBL (10)  & 7 ± 6 (10 ± 9)    & 373.3 ± 20.9    & 0.67 ± 0.09 & 2.08 ± 0.23 & 0.82 ± 0.17 & 1,000 (2.1h ± 24m)              \\[3pt]
Saturn-ZINC (9)    & 6 ± 3 (8 ± 10)     & 368.7 ± 27.6    & 0.79 ± 0.08 & 2.15 ± 0.22 & 0.87 ± 0.19 & 1,000 (2.1h ± 29m)               \\[3pt] 
Saturn-GA-ZINC (10)  & 7 ± 4 (10 ± 7)     & 382.8 ± 27.9    & 0.71 ± 0.08 & 2.10 ± 0.15 & 0.85 ± 0.17 & 1,000 (2.0h ± 26m)               \\[3pt] 
\midrule
\textbf{$R_{Double\ {MPO}}$} \textbf{(ours)} & \multicolumn{6}{c}{2 objectives (Docking, AiZynth)} \\
\midrule
Saturn-ChEMBL (10)   & 49 ± 19 (175 ± 94) & 442.3 ± 26.2    & 0.36 ± 0.05 & 2.36 ± 0.17 & 0.84 ± 0.06 & 1,000 (2.0h ± 31m)              \\[3pt]
Saturn-GA-ChEMBL (10)  & 43 ± 19 (99 ± 54)  & 436.1 ± 17.2    & 0.39 ± 0.04 & 2.35 ± 0.13 & 0.77 ± 0.07 & 1,000 (1.7h ± 16m)              \\[3pt]
Saturn-ZINC (10)    & 24 ± 17 (71 ± 64) & 414.0 ± 20.2    & 0.52 ± 0.11 & 2.30 ± 0.25 & 0.90 ± 0.07 & 1,000 (1.9h ± 22m)               \\[3pt] 
Saturn-GA-ZINC (10)    & 30 ± 11 (64 ± 28) & 408.1 ± 12.4    & 0.46 ± 0.05 & 2.19 ± 0.10 & 0.86 ± 0.07 & 1,000 (1.6h ± 16m)               \\[3pt] 
\bottomrule
\end{tabular}
\end{table}

In the previous section, we have shown that generative models can exploit docking oracles. Yet, docking scores can be valuable as they can be \textit{correlated} with better binding affinity~\cite{dockstream} and should be optimized in combination with oracles that modulate physico-chemical properties. In this section, we run Saturn with the $R_{All\ {MPO}}$ (jointly maximize QED, minimize SA score, minimize docking score, and is AiZynthFinder solvable) and $R_{Double\ {MPO}}$ (jointly minimize docking score and is AiZynthFinder solvable) objective functions.

\textbf{Quantitative Results.} Table \ref{table:synth-metrics} shows the Saturn results and also results taken from RGFN's~\cite{rgfn} work. \textbf{As stated previously, since the comparison to RGFN is not apples to apples, we focus our discussion on Saturn.} The central message of this section is that molecules satisfying the objective functions can be found within 1,000 oracle calls. An important note is that all RGFN results (top half of Table \ref{table:synth-metrics}) report results for the top-500 (for Mol. weight, QED~\cite{qed}, SA score~\cite{sa-score}) and top-100 (for AiZynthFinder~\cite{aizynthfinder-dataset, aizynthfinder, aizynthfinder-4}) Modes. Saturn does not find 100 Modes in all configurations with 1,000 oracle calls so the metrics are reported for however many Modes were found. Finally, we run Saturn with and without GraphGA-augmented experience replay~\cite{graphga, saturn} (see Appendix \ref{appendix:graphga-experience-replay} for details) and pre-trained with both ChEMBL 33~\cite{chembl} and ZINC 250k~\cite{zinc250k} (see Appendix \ref{appendix:pre-training} for pre-training details). The purpose is to show that the MPO task can be optimized in 1,000 oracle calls using both popular pre-training datasets. We make the following observations: by including AiZynthFinder in the objective function, Saturn generates AiZynthFinder solvable molecules. Including QED and SA score in the objective function also optimizes these metrics (contrast $R_{All\ {MPO}}$ with $R_{Double\ {MPO}}$ results). Mol. weight is also implicitly minimized because it is a component of QED. $R_{Double\ {MPO}}$ finds notably more Modes than $R_{All\ {MPO}}$ because the optimization task is easier. In all cases, 1/400th the oracle budget is sufficient to find at least \textit{some} molecules that optimize the objectives (and are AiZynthFinder solvable). The wall times are not 1/400th because AiZynthFinder is the slowest oracle, even with multi-threading (see Appendix \ref{appendix:aizynthfinder}). Finally, we highlight that although the raw number of Modes generated when using the $R_{All\ {MPO}}$ objective function is relatively low (in 1,000 oracle calls), if AiZynthFinder \textit{does} accurately predict "true" synthesizability, then these Modes are immediately actionable. Importantly, they satisfy every metric in the objective function (low docking score, high QED, low SA, and is AiZynthFinder solvable). In practice, one wants to identify a small set of \textit{excellent} candidate molecules as fast as possible (oracle calls and/or wall time). See Appendix \ref{appendix:supplementary-results} for additional experiments, and particularly how \textit{also} optimizing for QED is a considerably more difficult task.

\begin{figure}
\centering
\includegraphics[width=1.0\columnwidth]{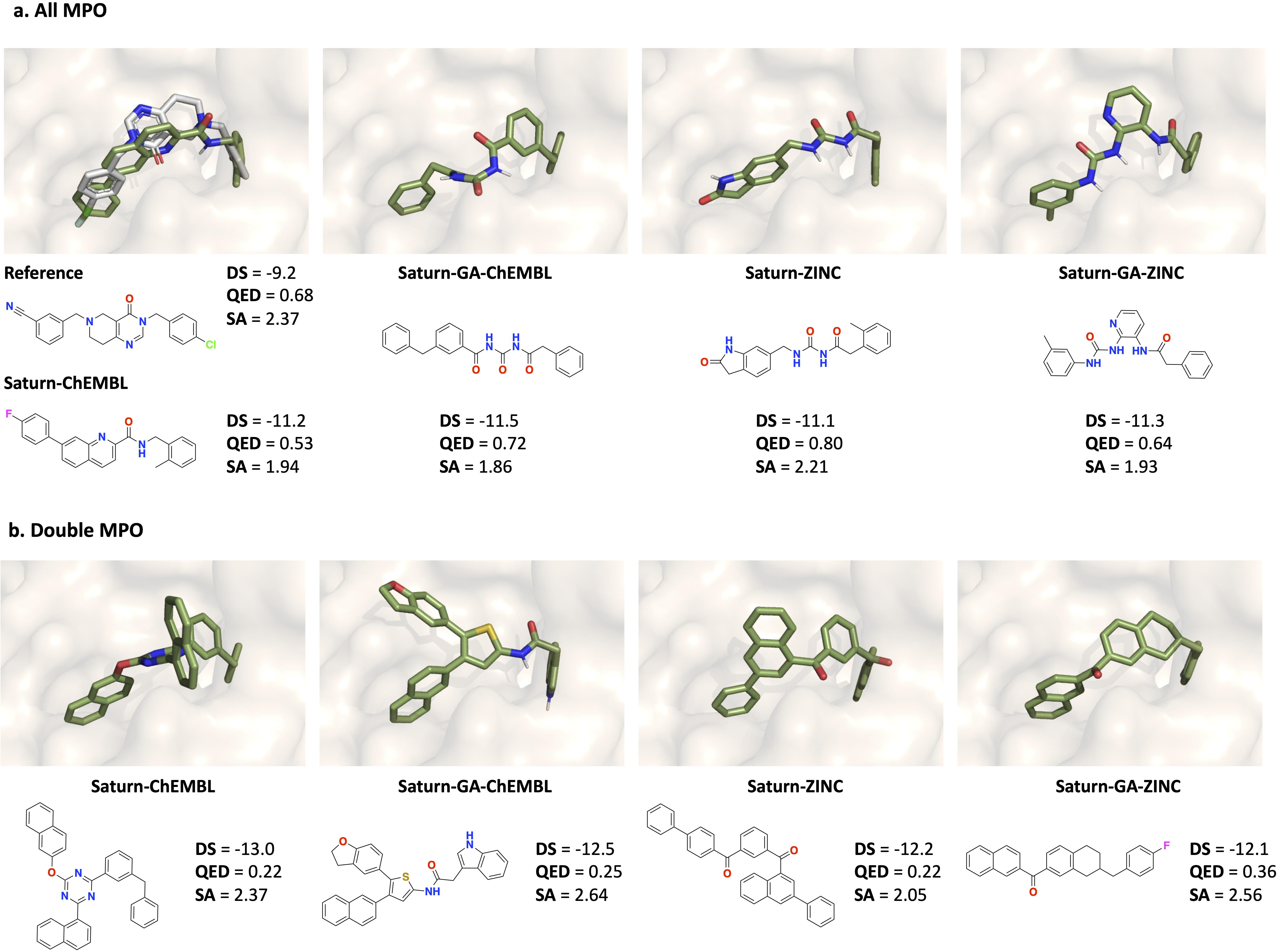} 
\caption{Docked pose of the reference ligand (PDB ID: 7UVU) and generated molecules with the best docking score (DS) across all Saturn configurations and across all 10 seeds (0-9 inclusive). The reference pose is in gray and all generated molecules are in green. \textbf{All molecules are AiZynthFinder solvable.} \textbf{a.} Molecules generated using $R_{All\ {MPO}}$. \textbf{b.} Molecules generated using $R_{Double\ {MPO}}$.}
\label{fig:example-poses}
\end{figure}

\textbf{Qualitative Results.} Fig. \ref{fig:example-poses} shows the docking pose for the generated molecules with the best docking score (no cherry-picking) across all Saturn configurations. In all cases, the pose conforms to the geometry of the binding cavity and the molecule itself is AiZynthFinder solvable (see Appendix \ref{appendix:aizynthfinder-routes} for the solved routes). Generated molecules using $R_{Double\ {MPO}}$ have better docking scores than $R_{All\ {MPO}}$, which is expected as the optimization task is easier. In the case of $R_{Double\ {MPO}}$, the best molecules have docking scores and QED values similar to the best molecules generated by RGFN~\cite{rgfn} in 400,000 oracle calls (Fig. \ref{fig:greasy}). We highlight that the molecules from $R_{Double\ {MPO}}$ possess extensive carbon rings and are likely exploiting the docking oracle.

\subsection{Experiment 3: Directly optimizing for synthesizability using AiZynthFinder starting from an unsuitable training distribution}

\begin{figure}
\centering
\includegraphics[width=1.0\columnwidth]{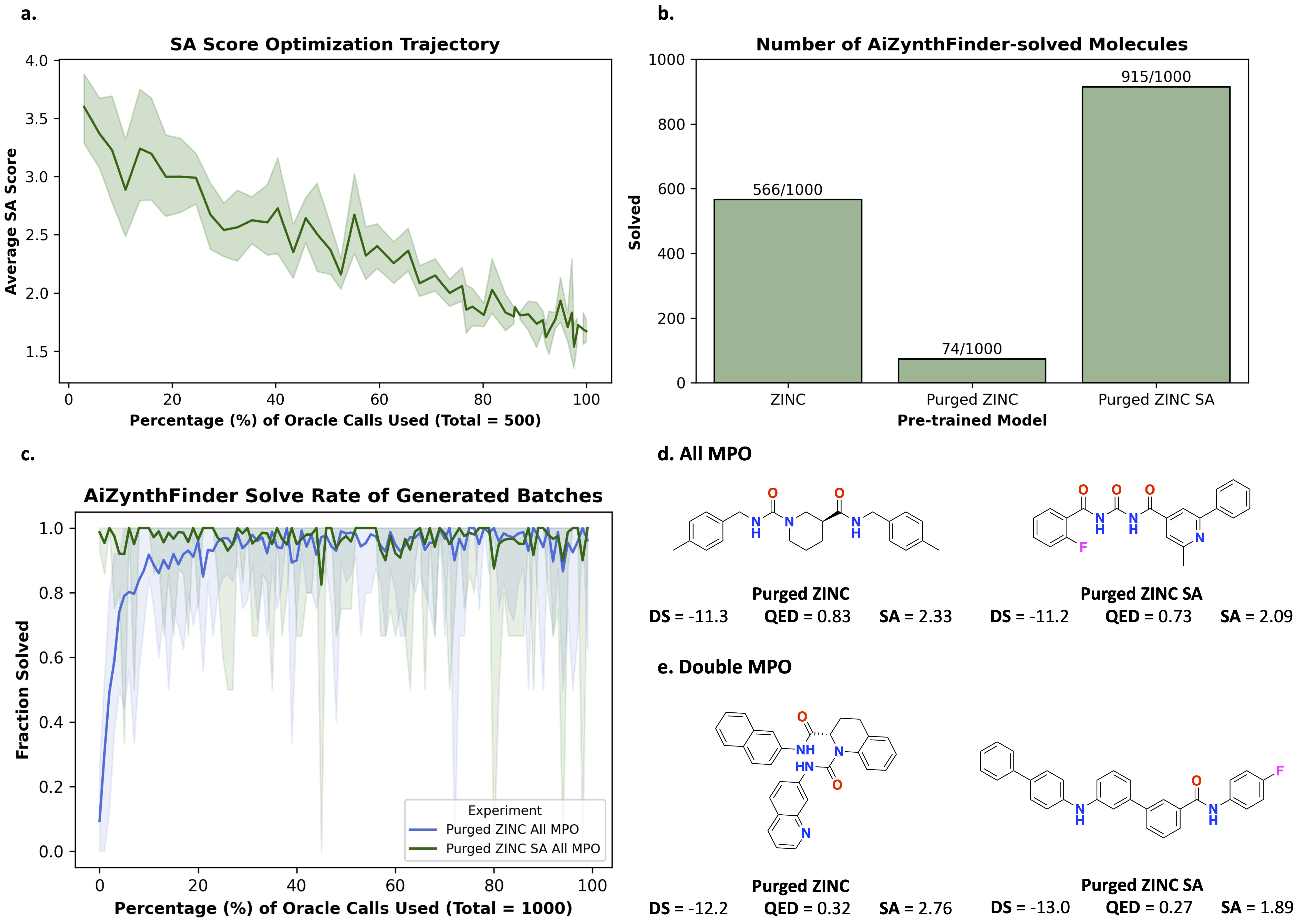} 
\caption{Correlation of SA score and AiZynthFinder solve rate and learning to generate AiZynthFinder solvable molecules. \textbf{a.} "Purged ZINC" is tasked to minimize SA score. The average SA score of the sampled batches are shown. \textbf{b.} AiZynthFinder solve rates for 1,000 molecules sampled from different models. \textbf{c.} All MPO task: fraction of generated molecules (without the GA activated) across all batches that are AiZynthFinder-solvable. Values are the mean and the shaded regions are the minimum-maximum across 10 seeds (0-9 inclusive). \textbf{d.} Example molecules generated from the "Purged ZINC" and "Purged ZINC SA" models with the best docking scores (DS).}
\label{fig:purge-recovery}
\end{figure}

Generative models are pre-trained to model the training data distribution. The experiments thus far use Saturn~\cite{saturn} which has been trained on either ChEMBL 33~\cite{chembl} or ZINC 250k~\cite{zinc250k}. These datasets contain bio-active molecules and pre-trained models can already generate molecules that can be solved by a retrosynthesis model~\cite{gao2020synthesizability}. In this section, we pre-train Saturn on the fraction of ZINC 250k that \textit{is not} AiZynthFinder~\cite{aizynthfinder-dataset, aizynthfinder, aizynthfinder-4} solvable. This model will be referred to as "Purged ZINC" (see Appendix \ref{appendix:purged-pre-training} for details). The message we convey is that even with an unsuitable training distribution, both $R_{All\ {MPO}}$ and $R_{Double\ {MPO}}$ can \textit{still} be optimized under a 1,000 oracle budget. 

We showcase how curriculum learning (CL)~\cite{curriculum-learning} (decompose a complex optimization objective into sequential, simpler objectives) can be used by defining two phases of goal-directed generation. Firstly, "Purged ZINC" is tasked to minimize SA score~\cite{sa-score} (500 oracle budget) as it is correlated with AiZynthFinder~\cite{synthesizability-metrics-assessment}. Fig. \ref{fig:purge-recovery}a shows the optimization trajectory and the resulting model is referred to as "Purged ZINC SA". The 500 oracle calls are not counted in the 1,000 oracle budget, as computing SA score is cheap (this process took 56 seconds). Next, to illustrate distribution learning, we sample 1,000 unique molecules from the "Normal ZINC" (trained on the full dataset), "Purged ZINC", and "Purged ZINC SA" models and run AiZynthFinder. Fig. \ref{fig:purge-recovery}b shows the fraction of molecules that are AiZynthFinder solvable. In 56 seconds, the "Purged ZINC" model can be fine-tuned to immediately generate molecules that are almost all solvable (see Appendix \ref{appendix:supplementary-results} for additional results and discussion). We note that "Purged ZINC" still generates molecules that are AiZynthFinder solvable due to stochastic generation and likely due to the use of SMILES randomization during training which enhances chemical space generalizability~\cite{arus2019randomized}. Next, we show how the "Purged ZINC" model can learn to generate molecules that are AiZynthFinder solvable during the course of RL (Fig. \ref{fig:purge-recovery}c). We contrast this with the "Purged ZINC SA" model which has almost 100\% solve rate throughout the entire run. During the course of the run, some seeds occasionally generate batches that are not AiZynthFinder solvable (lower bound of shaded region), but this is not detrimental (see Appendix \ref{saturn-batch-generation} for more details). Fig. \ref{fig:purge-recovery}d shows the molecules with the best docking score generated across all seeds. The property profiles are essentially the same as the runs with the normal ZINC model (Fig. \ref{fig:example-poses}).

\begin{table}[ht]
\centering
\scriptsize
\caption{Synthesizability metrics for "Normal ZINC" (results from Table \ref{table:synth-metrics}), "Purged ZINC", and "Purged ZINC SA". All experiments were run across 10 seeds (0-9 inclusive). The mean and standard deviation are reported. Both Yield and Modes are reported. The number after the configuration denotes the number of successful replicates out of 10 (Modes $\geq$ 1). The metrics are reported for however many Modes were found.}
\label{table:synth-metrics-purged}
\begin{tabular}{@{}lcccccc@{}}
\toprule
Method    & Modes (Yield) & Mol. weight (↓) & QED (↑) & SA score (↓) & AiZynth (↑) & Oracle calls \\ 
          &       &                 &         &             &             &          (Wall time) \\ 
\midrule
\textbf{$R_{All\ {MPO}}$} & \multicolumn{6}{c}{4 objectives (Docking, QED, SA, AiZynth)} \\
\midrule
\textbf{Normal ZINC} \\[3pt]
Saturn (9)    & 6 ± 3 (8 ± 10)     & 368.7 ± 27.6    & 0.79 ± 0.08 & 2.15 ± 0.22 & 0.87 ± 0.19 & 1,000 (2.1h ± 29m)               \\[3pt] 
Saturn-GA (10)  & 7 ± 4 (10 ± 7)    & 382.8 ± 27.9    & 0.71 ± 0.08 & 2.10 ± 0.15 & 0.85 ± 0.17 & 1,000 (2.0h ± 26m)   \\
\midrule
\textbf{Purged ZINC} \\[3pt]
Saturn (8)   & 5 ± 5 (9 ± 15)   & 354.4 ± 26.2    & 0.72 ± 0.15 & 1.99 ± 0.27 & 0.97 ± 0.05  & 1,000 (2.8h ± 72m)            \\[3pt]
Saturn-GA(10)  & 10 ± 3 (14 ± 5)   & 381.4 ± 15.6    & 0.68 ± 0.09 & 2.22 ± 0.24 & 0.77 ± 0.12 & 1,000 (2.4h ± 50m) \\
\midrule
\textbf{Purged ZINC SA} \\[3pt]
Saturn (10)    & 9 ± 5 (16 ± 11)     & 365.9 ± 12.5    & 0.68 ± 0.09 & 1.97 ± 0.19 & 0.96 ± 0.10 & 1,000 (4.9h ± 54m)               \\[3pt] 
Saturn-GA (10)  & 12 ± 6 (21 ± 14)    & 369.7 ± 15.0    & 0.69 ± 0.08 & 2.06 ± 0.15 & 0.89 ± 0.08 & 1,000 (3.2h ± 26m)  \\
\midrule
\textbf{$R_{Double\ {MPO}}$} & \multicolumn{6}{c}{2 objectives (Docking, AiZynth)} \\
\midrule 
\textbf{Normal ZINC} \\[3pt]
Saturn (10)    & 24 ± 17 (71 ± 64) & 414.0 ± 20.2    & 0.52 ± 0.11 & 2.30 ± 0.25 & 0.90 ± 0.07 & 1,000 (1.9h ± 22m)               \\[3pt] 
Saturn-GA (10)    & 30 ± 11 (64 ± 28) & 408.1 ± 12.4    & 0.46 ± 0.05 & 2.19 ± 0.10 & 0.86 ± 0.07 & 1,000 (1.6h ± 16m)               \\
\midrule
\textbf{Purged ZINC} \\[3pt]
Saturn (10)   & 27 ± 19 (114 ± 107) & 425.7 ± 58.5    & 0.50 ± 0.15 & 2.66 ± 0.56 & 0.83 ± 0.13 & 1,000 (2.4h ± 35m)              \\[3pt]
Saturn-GA (10) & 34 ± 17 (78 ± 57) & 410.8 ± 16.3    & 0.44 ± 0.08 & 2.29 ± 0.16 & 0.78 ± 0.08 & 1,000 (2.2h ± 34m)               \\
\midrule
\textbf{Purged ZINC SA} \\[3pt]
Saturn (10)    & 46 ± 14 (268 ± 88) & 443.5 ± 31.4    & 0.39 ± 0.10 & 2.13 ± 0.12 & 0.87 ± 0.08 & 1,000 (3.9h ± 56m)               \\[3pt] 
Saturn-GA (10) & 49 ± 10 (187 ± 55) & 419.5 ± 10.6    & 0.42 ± 0.04 & 2.11 ± 0.05 & 0.77 ± 0.06 & 1,000 (2.9h ± 29m)               \\
\bottomrule
\end{tabular}
\end{table}

\textbf{Quantitative Results.} Table \ref{table:synth-metrics-purged} contrasts the results of the ClpP docking case study run across 10 seeds (0-9 inclusive) using the "Normal ZINC", "Purged ZINC", and "Purged ZINC SA" models. Despite an unsuitable training distribution, "Purged ZINC" can still generate Modes that are AiZynthFinder solvable, although the solve rate is slightly lower than "Normal ZINC". "Purged ZINC SA" was first fine-tuned to minimize SA score and already generated mostly AiZynthFinder solvable molecules (Fig. \ref{fig:purge-recovery}b). This process benefits both $R_{All\ {MPO}}$ (less so) and $R_{Double\ {MPO}}$ as the Yield and Modes found are higher. Next, we highlight that "Purged ZINC" and "Purged ZINC SA" wall times are longer. This is due to two reasons: firstly, we ran four experiments simultaneously on a single workstation, which shares resources but makes the total wall time to finish all the experiments faster. Secondly, "Purged ZINC SA" experiments took longer because the initial CL fine-tuning biases the model to generate more repeat molecules due to Saturn's~\cite{saturn} mechanism of local chemical space exploration. The effect is that it takes longer to exhaust the 1,000 \textit{unique} oracle calls budget.

Overall, the property profiles of generated Modes are better than GraphGA~\cite{graphga}, SyntheMol~\cite{synthemol}, FGFN~\cite{bengio2021flow}, and RGFN~\cite{rgfn}. Regardless of the starting model, both $R_{All\ {MPO}}$ and $R_{Double\ {MPO}}$ can be optimized within 1,000 oracle calls. Whether or not the output of AiZynthFinder represents "true" synthesizability (and the quality of the routes) is beyond the scope of this work. The message we convey in this section is that generating molecules that are solvable by a retrosynthesis model does not \textit{require} synthesizability-constrained design principles. Lastly, we explore the effect of increasing the oracle budget and implications of heuristics-driven synthesizability and post-hoc retrosynthesis model filtering in Appendix \ref{appendix:supplementary-results}.

\section{Conclusion}

In this work, we adapt Saturn~\cite{saturn} which is a sample-efficient autoregressive molecular generative model using the Mamba~\cite{mamba} architecture to directly optimize for synthesizability using retrosynthesis models~\cite{ekborg2024novo}. Our approach contrasts existing works in the field that tackle synthesizability in one of three ways: goal-directed generation with synthesizability heuristic scores such as SA score~\cite{sa-score}, post-hoc filtering generated molecules with a retrosynthesis model~\cite{az-aizynthfinder}, or by enforcing synthesizability design principles in the generative process itself (synthesizability-constrained generation)~\cite{synnet, synflownet, rgfn}. We show that with a sufficiently sample-efficient model, treating retrosynthesis models as an oracle is feasible, and generated molecules can satisfy multi-parameter optimization objectives while being synthesizable (as deemed by a retrosynthesis model). The main comparison results we show are on a molecular docking case study proposed by the recent Reaction-GFlowNet (RGFN)~\cite{rgfn} work which is a synthesizability-constrained generative model. With 1/400th the oracle budget, Saturn can generate molecules with better property profiles than GraphGA~\cite{graphga}, SyntheMol~\cite{synthemol}, Fragment GFlowNet (FGFN)~\cite{bengio2021flow}, and RGFN. Moreover, we conduct an artificial experiment to intentionally purge a training dataset of all molecules that are solvable by the AiZynthFinder~\cite{aizynthfinder-dataset, aizynthfinder, aizynthfinder-4} retrosynthesis model and pre-train a new model with this dataset. Generated molecules from this model are mostly not AiZynthFinder solvable, as expected (Fig. \ref{fig:purge-recovery}b). Despite this, we show that within 1/400th the oracle budget, this model can \textit{still} generate molecules with property profiles better than all comparing models and are synthesizable (as deemed by AiZynthFinder). The take-home message is that with a sufficiently sample-efficient model, it is straightforward to treat retrosynthesis models as an oracle in goal-directed generation. Generating molecules deemed synthesizable by such models does not \textit{require} synthesizability-constrained generation, which is currently, often \textit{sample-inefficient}.

\newpage
\citestyle{nature}
\bibliographystyle{unsrtnat}  
\bibliography{bibliography}

\newpage
\appendix
\renewcommand\thefigure{\thesection\arabic{figure}}

\section*{Acknowledgement}

We thank Rebecca M. Neeser and Zlatko Jončev for their feedback on the draft. This publication was created as part of NCCR Catalysis (grant number 180544), a National Centre of Competence in Research funded by the Swiss National Science Foundation.

\section*{Appendix}
The Appendix contains details on the procedure we took to reproduce RGFN's~\cite{rgfn} oracle as the code is not released. In addition, we report the computational resources used, how Saturn was pre-trained, and AiZynthFinder execution details. The code is available at \url{https://github.com/schwallergroup/saturn/tree/synth}.

\section{Compute Resources}
\label{appendix:compute-resources}

All experiments were run on a single workstation with an NVIDIA RTX A6000 GPU 48GB memory and AMD Ryzen 9 5900X 24-Core CPU. 48GB GPU memory is not required. QuickVina2-GPU-2.1~\cite{autodockvina, quickvina2, quickvina2-gpu-2.1} with `thread` = 8,000 (following the RGFN~\cite{rgfn} work) takes up to 12GB GPU memory. We further note that Saturn's wall times reported in Table \ref{table:synth-metrics} are longer than actually required as we always run 2-4 experiments in parallel, which share the workstation's resources, but makes the \textit{total} wall time less.

\section{Saturn Pre-training Details}
\label{appendix:pre-training}

This section contains the exact protocol used for Saturn pre-training on ChEMBL 33~\cite{chembl} and ZINC 250k~\cite{zinc250k}. The details and pre-trained models are taken from the original Saturn~\cite{saturn} paper and included here.

\subsection{ChEMBL 33}

Each step is followed by the SMILES remaining after the filtering step.

\begin{enumerate}
    \item{Download raw ChEMBL 33 - 2,372,674}
    \item{Standardization (charge and isotope handling) based on \url{https://github.com/MolecularAI/ReinventCommunity/blob/master/notebooks/Data_Preparation.ipynb}. All SMILES that could not be parsed by RDKit were removed - 2,312,459}
    \item{Kept only the unique SMILES - 2,203,884}
    \item{Tokenize all SMILES based on REINVENT's tokenizer: \url{https://github.com/MolecularAI/reinvent-models/blob/main/reinvent_models/reinvent_core/models/vocabulary.py}}
    \item{Keep SMILES $\le$ 80 tokens - 2,065,099}
    \item{150 $\le$ molecular weight $\le$ 600 - 2,016,970}
    \item{Number of heavy atoms $\le$ 40 - 1,975,282}
    \item{Number of rings $\le$ 8 - 1,974,522}
    \item{Size of largest ring $\le$ 8 - 1,961,690}
    \item{Longest aliphatic carbon chain $\le$ 5 - 1,950,213}
    \item{Removed SMILES containing the following tokens (due to undesired chemistry and low token frequency): [S+], [C-], [s+], [O], [S@+], [S@@+], [S-], [o+], [NH+], [n-], [N@], [N@@], [N@+], [N@@+], [S@@], [C+], [S@], [c+], [NH2+], [SH], [NH-], [cH-], [O+], [c-], [CH], [SH+], [CH2-], [OH+], [nH+], [SH2] - \textbf{1,942,081}}
\end{enumerate}

The final vocabulary contained 37 tokens (2 extra tokens were added, indicating <START> and <END>). 

The Mamba model has 5,265,920 parameters. The hyperparameters are the default parameters in the code base.

\textbf{The pre-training parameters were}:

\begin{enumerate}
    \item{Max training steps = 20 (each training step entails a full pass through the dataset)}
    \item{Seed = 0}
    \item{Batch size = 512}
    \item{Learning rate = 0.0001}
    \item{Randomize~\cite{smiles-enumeration} every batch of SMILES}
\end{enumerate}

The following checkpoint was used: Epoch 18, NLL = 32.21, Validity (10k) = 95.60\%.

\subsection{ZINC 250k}

ZINC 250k~\cite{zinc250k} was downloaded and used as is.

\textbf{The pre-training parameters were}:

\begin{enumerate}
    \item{Training steps = 50 (each training step entails a full pass through the dataset)}
    \item{Seed = 0}
    \item{Batch size = 512}
    \item{Learning rate = 0.0001}
    \item{Train with SMILES randomization~\cite{smiles-enumeration} (all SMILES in each batch was randomized)}
\end{enumerate}

The final vocabulary contained 66 tokens (2 extra tokens were added, indicating <START> and <END>). 

The Mamba model has 5,272,832 parameters (slightly larger than ChEMBL 33 model because the vocabulary size here is larger). The following checkpoint was used: Epoch 50, NLL = 28.10, Validity (10k) = 95.20\%.

\section{Reproducing RGFN's Oracle}
\label{appendix:rgfn-oracle}

This section contains the steps we took to reproduce RGFN's~\cite{rgfn} ATP-dependent Clp protease proteolytic subunit (ClpP) docking case study as faithfully as we could. 

\textbf{Target Preparation.} Following Appendix C.1 of the RGFN paper, we downloaded the 7UVU ClpP crystal structure here: \url{https://www.rcsb.org/structure/7UVU}. All molecules (complexed inhibitors, solvents, etc.) were removed, keeping only two monomeric units. Two structures were saved: The apo protein (no other molecules present) and the reference ligand. \textbf{The following step differs from RGFN}: the apo protein was processed with PDBFixer~\cite{pdbfixer} to fix missing atoms and residues. We performed this step because errors were thrown during docking when using the raw apo protein structure.

\textbf{Docking Details.} We implement QuickVina2-GPU-2.1~\cite{autodockvina, quickvina2, quickvina2-gpu-2.1} following the instructions in the GitHub repository here: \url{https://github.com/DeltaGroupNJUPT/Vina-GPU-2.1}. The reference ligand structure that was saved out in the previous step is used here to define the docking box. Specifically, the average coordinates of the ligand denote the docking centroid. \textbf{The following \textit{may} differ from RGFN}: We define the docking box as 20 Å x 20 Å x 20 Å as it was unclear how it should be defined based on RGFN's protocol. This box size has worked on many other protein targets~\cite{dockstream} when docking with AutoDock Vina~\cite{autodockvina} which is the predecessor of QuickVina2-GPU-2.1.

\textbf{Docking Workflow.} Following RGFN's protocol, QuickVina2-GPU-2.1 used the following parameters: `thread` = 8,000 with `search depth` = "heuristic" which is the default. Next, all ligands were docked following RGFN's workflow:

\begin{enumerate}
    \item{Start with batch of generated SMILES from Saturn}
    \item{Canonicalize the SMILES}
    \item{Convert to RDKit Mol objects}
    \item{Protonate the Mols}
    \item{Generate 1 (lowest energy) conformer using `ETKDG`~\cite{etkdg}}
    \item{Minimize energy with the Universal Force Field (UFF)~\cite{uff}}
    \item{Write out the conformers as `PDB` files}
    \item{Using Open Babel~\cite{obabel}, convert the `PDB` to `PDBQT` format}
    \item{Execute QuickVina2-GPU-2.1 docking}
\end{enumerate}

\textbf{Protocol Validation.} We make further efforts to ensure the oracle is as faithful as possible to RGFN's implementation. When executing QuickVina2-GPU-2.1, if a seed is not specified, a random seed is used. It is unclear if a seed was set in the RGFN~\cite{rgfn} work. In our experiments, the seed is 0. We re-dock the reference ligand and find that the pose is similar to Figure 15 in the RGFN work. However, the docking score we obtain is -9.2 whereas the RGFN work reports -10.31. Subsequently, we execute docking 100 times (letting QuickVina2-GPU 2.1 select the random seed) and observed that seed = 448029751 gives a similar pose to RGFN's pose and yields a docking score of -10.1. We additionally found that seed = 1920393356 yields a docking score of -10.3 but the pose is reflected. Finally seed = 673697018 yields a docking score of -8.2 and is a completely different pose. It is intractable to try every seed.

Therefore, we end this section by stating that it is hard to say if we \textit{exactly} re-implement RGFN's~\cite{rgfn} docking oracle. However, we believe it still enables us to convey the primary message of our work: retrosynthesis models can be directly treated as an oracle and be explicitly optimized for during generation.

\section{AiZynthFinder}
\label{appendix:aizynthfinder}

AiZynthFinder~\cite{aizynthfinder-dataset, aizynthfinder, aizynthfinder-4} was used as is, without modification. The source code was cloned from the GitHub repository here: \url{https://github.com/MolecularAI/aizynthfinder}. The environment and package were installed following the README. Following the documentation here: \url{https://molecularai.github.io/aizynthfinder/}, we downloaded the public data and used AiZynthFinder as is. Every batch of molecules generated by Saturn (16 at max) is chunked into 4 sub-sets for multi-thread execution. Finally, we consider a molecule AiZynthFinder "solvable" if the "is\_solved" flag is True. This flag denotes whether the top scored (accounting for tree depth and fraction of building blocks in stock)~\cite{aizynthfinder, aizynthfinder-4} is solved.

\section{AiZynthFinder purged ZINC 250k Pre-training Details.}
\label{appendix:purged-pre-training}

In Experiment 3, we pre-train Saturn on a sub-set of ZINC 250k~\cite{zinc250k} that \textit{is not} AiZynthFinder~\cite{aizynthfinder-dataset, aizynthfinder, aizynthfinder-4} solvable. The goal is to show that Saturn can \textit{still} optimize for generating molecules that are AiZynthFinder solvable despite being trained on no molecules that can be.

\textbf{Purged Dataset.} We first run AiZynthFinder on the entirety of ZINC 250k on a single workstation with an NVIDIA RTX A6000 GPU 48GB memory and AMD Ryzen 9 5900X 24-Core CPU. The process was run using multi-threading across 12 workers and took 62 hours. We save the unique SMILES (98,110) of all the molecules that \textit{are not} AiZynthFinder solvable. This is the dataset used for pre-training.

\textbf{Pre-training.} Following the same pre-training parameters used in the original Saturn~\cite{saturn} work:

\begin{enumerate}
    \item{Training steps = 100 (each training step entails a full pass through the dataset)}
    \item{Seed = 0}
    \item{Batch size = 512}
    \item{Learning rate = 0.0001}
    \item{Train with SMILES randomization~\cite{smiles-enumeration} (all SMILES in each batch was randomized)}
\end{enumerate}

The final vocabulary contained 57 tokens (2 extra tokens were added, indicating <START> and <END>).  This is less than the normal ZINC 250k model (66 tokens) because some tokens are not present in the purged dataset.

The Mamba model has 5,271,040 parameters (less than the normal 250k model because the vocabulary size is smaller). The following checkpoint was used: Epoch 100, NLL = 27.78, Validity (10k) = 92.27\% and the training time was 4.7 hours.

\section{AiZynthFinder Routes}
\label{appendix:aizynthfinder-routes}

\begin{figure}
\centering
\includegraphics[width=1.0\columnwidth]{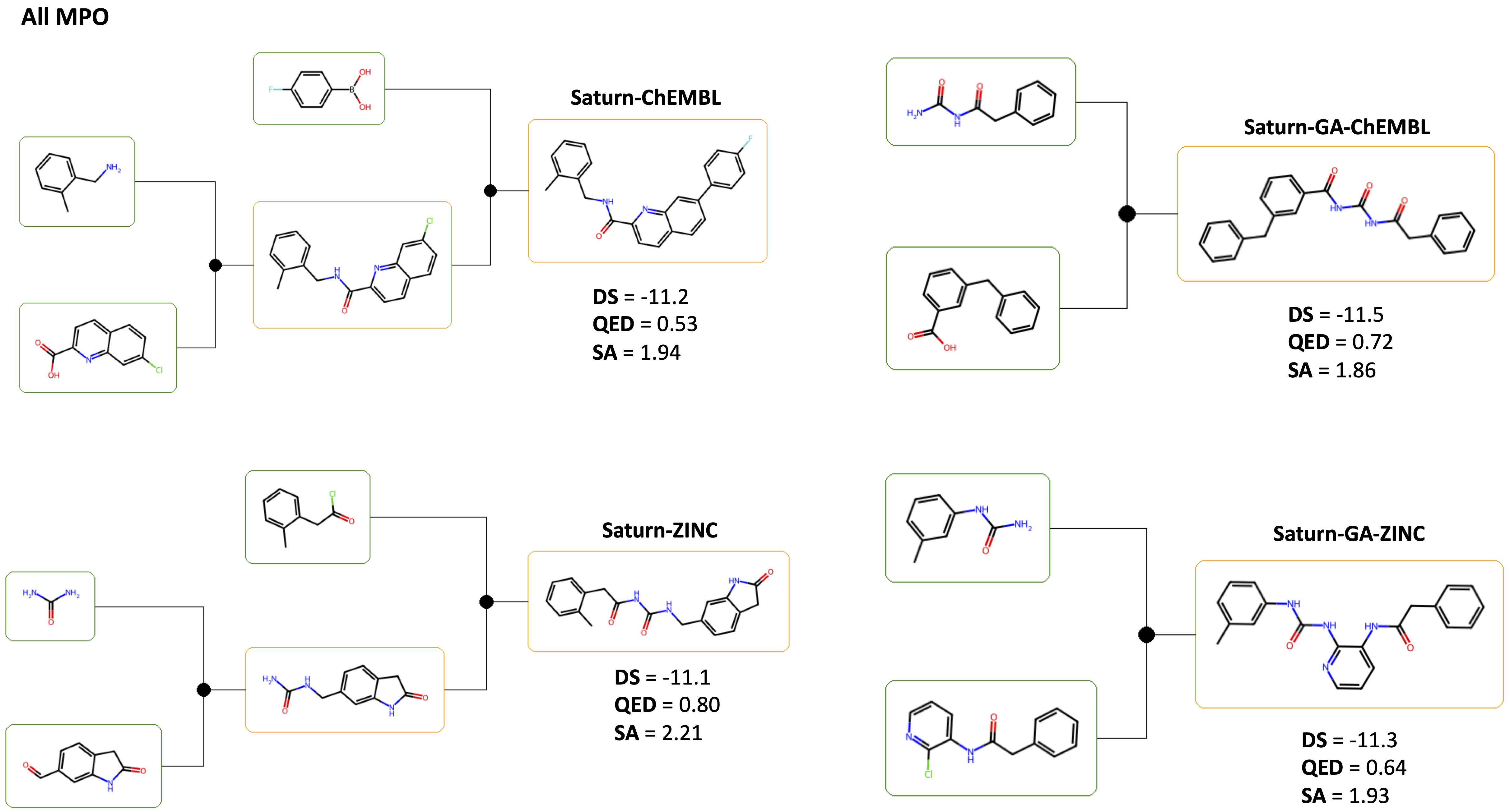}
\caption{AiZynthFinder solved routes (top-scoring) for All MPO example molecules.}
\label{fig:all-mpo-routes}
\end{figure}

\begin{figure}
\centering
\includegraphics[width=1.0\columnwidth]{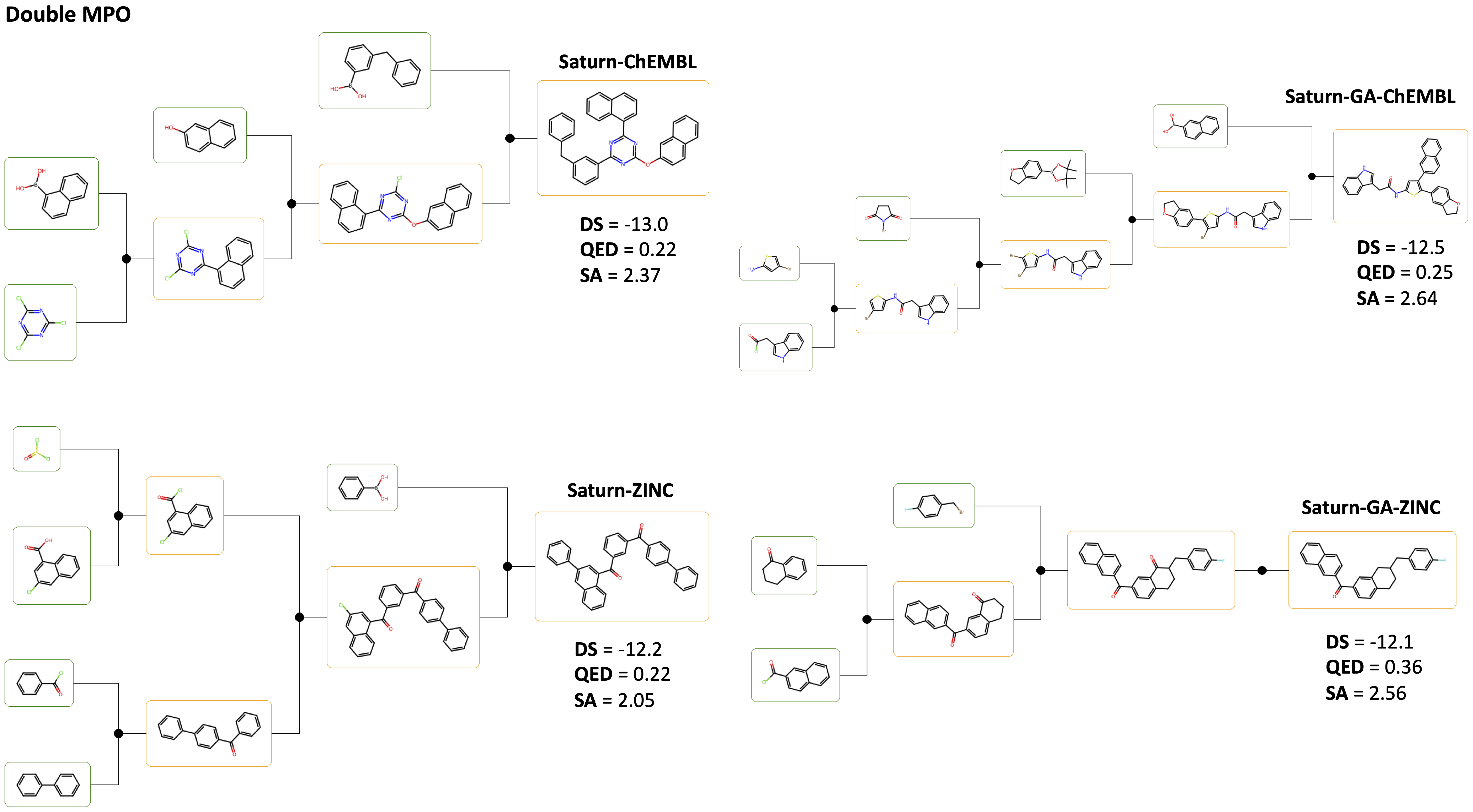} 
\caption{AiZynthFinder solved routes (top-scoring) for Double MPO example molecules.}
\label{fig:double-mpo-routes}
\end{figure}

The AiZynthFinder solved routes for the 8 example molecules shown in Fig. \ref{fig:example-poses} are shown here. The All MPO routes (Fig. \ref{fig:all-mpo-routes}) are generally shorter than the Double MPO routes (Fig. \ref{fig:double-mpo-routes}). This suggests that enforcing QED and SA score also implicitly makes the predicted forward syntheses shorter. We note that it is possible to design an objective function that also aims to generate short paths by rewarding short paths. We do not explore this here and leave it for future work.

\section{Supplementary Results}
\label{appendix:supplementary-results}

In this section, supplementary results are reported which aim to address/provide evidence for three points:

\begin{enumerate}
    \item{Effect of increasing the oracle budget when optimizing AiZynthFinder}
    \item{\textit{Jointly} optimizing QED with docking score is \textit{considerably} more difficult than just optimizing docking score}
    \item{Optimizing SA score \textit{can} be a better allocation of computational resources}
\end{enumerate}

\begin{table}[ht]
\centering
\scriptsize
\caption{Synthesizability metrics across various Saturn experiments. Metrics are reported for however many Modes are found. For these supplemental results, only one replicate is performed with seed = 0.}
\label{table:supplementary-results}
\begin{tabular}{@{}lcccccc@{}}
\toprule
Method    & Modes (Yield) & Mol. weight (↓) & QED (↑) & SA score (↓) & AiZynth (↑) & Oracle calls \\ 
          &       &                 &         &             &             &          (Wall time) \\ 
\midrule
\textbf{$R_{All\ {MPO}}$} & \multicolumn{6}{c}{4 objectives (Docking, QED, SA, AiZynth)} \\
\midrule
Saturn-GA-ChEMBL & 108 (222) & 370.9   & 0.84 & 2.44 & 0.70 & 5,000 (23.6h)               \\[3pt] 
Saturn-GA-ZINC  & 74 (230) & 371.1   & 0.81 & 2.45 & 0.69 & 5,000 (21.2h)   \\
\midrule
\textbf{$R_{Double\ {MPO}}$} & \multicolumn{6}{c}{2 objectives (Docking, AiZynth)} \\
\midrule 
Saturn-ChEMBL & 302 (3804) & 486.2   & 0.28 & 2.40 & 0.82 & 5,000 (21.4h)            \\[3pt] 
Saturn-GA-ChEMBL  & 323 (3053) & 464.1   & 0.34 & 2.51 & 0.68 & 5,000 (11.3h)                 \\[3pt] 
Saturn-ZINC & 266 (2783) & 521.2  & 0.25 & 2.40 & 0.76 & 5,000 (13.5h)               \\[3pt] 
Saturn-GA-ZINC  & 327 (2741) & 455.6  & 0.34 & 2.48 & 0.72 & 5,000 (11.3h)   \\
\midrule
\textbf{$R_{All\ {MPO}}$} (but without AiZynth) & \multicolumn{6}{c}{3 objectives (Docking, QED, SA)} \\
\midrule 
Saturn-ChEMBL & 332 (1219) & 376.7   & 0.80 & 2.66 & 0.39 & 10,000 (2.4h)               \\[3pt] 
Saturn-ZINC  & 332 (1108) & 382.1   & 0.76 & 2.43 & 0.55 & 10,000 (2.3h)   \\
\midrule
\textbf{$R_{RGFN}$} - Results from Fig. \ref{fig:greasy} & \multicolumn{6}{c}{1 objective (Docking)} \\
\midrule 
Saturn-ChEMBL  & 469 (8389) & 511.9 & 0.26 & 3.09 & 0.14 & 10,000 (2.2h)   \\
\bottomrule
\end{tabular}
\end{table}

\textbf{Increasing the oracle budget leads to notably increased wall times.} In the main text results, $R_{All\ {MPO}}$ does not find that many Modes. We investigate the effect of increasing the oracle budget (Table \ref{table:supplementary-results}) with the GA activated (which recover diversity so as to satisfy the Modes criterion that Modes must have < 0.5 Tanimoto similarity with other Modes). More Modes are found but the wall time is \textit{drastically} higher. With 5x the oracle budget (5,000 compared to 1,000 in the main text), one may expect 5x the wall time (12-15 hours) but the wall time is almost 24 hours. The reason is due to Saturn's sampling behaviour which locally explores chemical space~\cite{saturn}. The parameters of Saturn could be changed to loosen this local exploration behaviour but we do not explore this. We demonstrate the application of Saturn out-of-the-box. As a consequence of this, many repeat molecules are generated, which do not impose an oracle call as the reward is retrieved from an oracle cache, but makes the sampled batch (new molecules) smaller. Multi-threading was used to run AiZynthFinder faster (see Appendix \ref{appendix:aizynthfinder} for more details). Consider batches of 1 molecule and 4 molecules. This can take a similar wall time as molecules can be chunked, thus benefiting from multi-threading. This could be mitigated, for example, by using a faster retrosynthesis model which can come with advantages and disadvantages~\cite{retro*, syntheseus} and/or CPU parallelization. Finally, we highlight that deactivating the GA will likely lead to higher Yield and AiZynthFinder solve rate, as shown in Tables \ref{table:synth-metrics} and \ref{table:synth-metrics-purged}. We reiterate that activating the GA was to satisfy the Mode metric.

\textbf{\textit{Jointly} optimizing QED with docking score is \textit{considerably} more difficult than just optimizing docking score.} RGFN~\cite{rgfn} reports their mean and standard deviation of QED values as 0.23 ± 0.04 (unclear how many replicates this was over). This is low and suggests the model is exploiting the docking algorithm as shown in Fig. \ref{fig:greasy}. To show that \textit{jointly} optimizing QED and docking is a \textit{considerably} more difficult task, we first cross-reference the results for $R_{Double\ {MPO}}$ (Table \ref{table:supplementary-results}) where the Modes and Yield are notably higher than $R_{All\ {MPO}}$. Next, we cross-reference the results when QED is \textit{not} being optimized (Table \ref{table:supplementary-results} last row). The Yield is \textit{much} higher (molecules with docking score < 10,000) but the QED values are similar to RGFN, which again, suggests the docking algorithm is being exploited.

\textbf{Optimizing SA score \textit{can} be a better allocation of computational resources.} SA score~\cite{sa-score} is correlated with AiZynthFinder solve rate~\cite{synthesizability-metrics-assessment}. In the main text Fig \ref{fig:purge-recovery}, we empirically demonstrate this, as 56 seconds of fine-tuning a pre-trained model that has \textit{never} seen an AiZynthFinder solved molecule, results in a model that generates molecules almost all solvable. The natural next question is, would simply optimizing SA score be a better allocation of computational resources (as is commonly done)? Under the same wall time, many more queries to SA score can be made because it is computationally cheap. Correspondingly, we use the $R_{All\ {MPO}}$ objective function but omit AiZynthFinder (only docking, QED, and SA score) and run the ChEMBL and ZINC pre-trained models for 10,000 oracle calls (Table \ref{table:supplementary-results}). Firstly, the wall time is similar to running 1,000 oracle calls of AiZynthFinder (cross-reference Table \ref{table:synth-metrics}). Next, while a smaller fraction of the Modes are AiZynthFinder solvable, the raw number is higher than directly optimizing AiZynthFinder. This reinforces that post-hoc retrosynthesis model filtering is valid and is often what is done in practice~\cite{az-aizynthfinder}. Crucially, the actual percentage of AiZynthFinder solve rate may not \textit{actually} matter. What matters is that a user can reasonably expect a generative model to generate molecules satisfying the objective function within the allotted oracle budget and/or wall time. In this specific example, it does not matter that Saturn-ZINC "only" has 55\% solve rate when optimizing docking, QED, and SA score (Table \ref{table:supplementary-results}). Running the 332 Modes through AiZynthFinder only took about 20 minutes (about 183/332 can be solved). A user would only care that in under 3 hours, 183 Modes were found that have low docking score, high QED, low SA score, and are AiZynthFinder solvable.

Finally, we wish to be prudent with making definitive statements about whether just optimizing SA score is strictly \textit{better} than including a retrosynthesis model in the objective function. In this section alone, we have highlighted that different retrosynthesis models can have a large impact on wall time~\cite{retro*, syntheseus}, where faster wall times would narrow the gap between SA score's wall time. Moreover, molecules deemed difficult to synthesize by SA score may actually be straightforward to synthesize. Retrosynthesis models have much more flexibility as the building block stock and reactions can be changed, whereas SA score was designed based on the fixed PubChem corpus~\cite{sa-score}. One could even constrain the retrosynthesis model to only include building blocks and reactions that are available in-house, similar to what was done in a collaborative work involving Pfizer~\cite{generate-what-you-can-make}. Thus, we leave a more thorough investigation regarding SA score optimization compared to various retrosynthesis models and search algorithms for future work. In this work, only the AiZynthFinder~\cite{aizynthfinder-dataset, aizynthfinder, aizynthfinder-4} retrosynthesis model was used, which leverages Monte Carlo Tree Search and ZINC building blocks.

\section{Saturn Reward Shaping}
\label{appendix:reward-shaping}

\begin{figure}
\centering
\includegraphics[width=1.0\columnwidth]{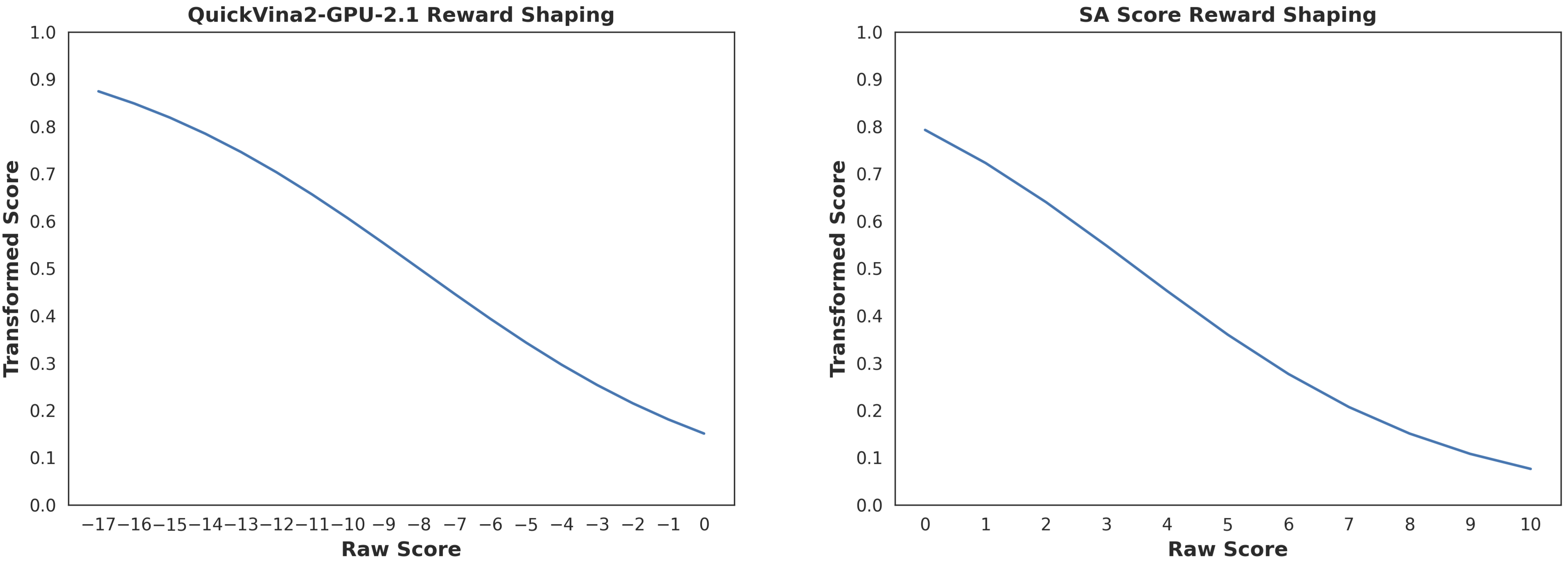}
\caption{Saturn reward shaping functions for QuickVina2-GPU-2.1 and SA score.}
\label{fig:reward-shaping}
\end{figure}

This section contains details on the reward shaping functions used such that the objective functions: $R_{RGFN}, R_{All\ {MPO}}, R_{Double\ {MPO}} \in [0, 1]$. Fig. \ref{fig:reward-shaping} shows the functions for QuickVina2-GPU-2.1~\cite{autodockvina, quickvina2, quickvina2-gpu-2.1} and SA score~\cite{sa-score}. QED~\cite{qed} values were taken as is, and not subjected to reward shaping. AiZynthFinder~\cite{aizynthfinder-dataset, aizynthfinder, aizynthfinder-4} returns 0 for not solved and 1 for solved. Given a molecule, all oracle evaluations are aggregated via a weighted product and a single scalar value is returned as the reward:

\begin{equation}
R(x) = \left[ \prod_i p_i(x)^{w_i} \right]^{\frac{1}{\sum_i w_i}}
\label{eq:weighted-product}
\end{equation}

$x$ is a SMILES~\cite{smiles}, $i$ is the index of an oracle given many oracles (MPO objective), $p_i$ is an oracle, and $w_i$ is the weight assigned to the oracle (1 for all oracles in this work).

\section{GraphGA-augmented Experience Replay}
\label{appendix:graphga-experience-replay}

Saturn~\cite{saturn} uses experience replay to enhance sample efficiency. GraphGA~\cite{graphga} can be applied on the replay buffer (stores the highest rewarding molecules generated so far) by treating the replay buffer as the parent population. Crossover and mutation operations then generate new molecules. For all the results in this work, activating the GA decreases the AiZynthFinder solve rate relative to no GA. This is because the generated molecules are not being sampled from the model itself (which is \textit{learning} to generate AiZynthFinder solvable molecules). What is gained in return is diversity recovering (as found in the original Saturn~\cite{saturn} work). This can be advantageous since the RGFN~\cite{rgfn} work defines \textbf{Discovered Modes} as the number of Modes (<-10 docking score) which also have < 0.5 Tanimoto similarity to every other mode. By activating the GA, more Modes are generally found, relative to no GA.

\section{Saturn Batch Generation}
\label{saturn-batch-generation}

Saturn~\cite{saturn} generates SMILES~\cite{smiles} in batches of, at maximum, 16. Internally, there is an oracle caching mechanism such that repeat generated SMILES are not sent for oracle evaluation, and instead, the reward is retrieved from the cache. Saturn's sample efficiency comes from the local exploration of chemical space, such that, at adjacent epochs, identical SMILES can be generated. The effect is that at each generation epoch, sometimes only a few \textit{new} (not generated before) SMILES are generated. In Fig. \ref{fig:purge-recovery}c, some batches have 0\% solve rate by AiZynthFinder. These are batches that only have a few new SMILES that happen not to be solvable. If one new SMILES is generated, it being unsolvable equates to 0\% solve rate.

\end{document}